\documentclass[doublecol]{epl2} 

\title{Magnetic Order Beyond RKKY in the Classical Kondo Lattice}
\shorttitle{Kondo Lattice Beyond RKKY} 

\author{Kalpataru Pradhan and Pinaki Majumdar}
\shortauthor{Kalpataru Pradhan and Pinaki Majumdar}

\institute{                    
\inst{} Harish-Chandra  Research Institute,
Chhatnag Road, Jhusi, Allahabad 211019, India 
}


\abstract{
We study the Kondo lattice model of band electrons coupled to
classical spins, in three dimensions, using a combination of
variational calculation and Monte Carlo.
We use the weak coupling `RKKY' window and the strong coupling
regime  as benchmarks, but focus on the physically relevant
intermediate coupling regime. Even for modest electron-spin
coupling the phase boundaries move away from the RKKY results,
the non interacting Fermi surface no longer dictates magnetic
order,
and  weak coupling `spiral' phases give way to collinear 
order. We use these results to revisit the classic problem of $4f$ 
magnetism and demonstrate how both electronic structure and 
coupling effects beyond RKKY control the magnetism in these 
materials.
}

\begin{document}

\maketitle

The Kondo lattice model describes local moments on a 
lattice coupled to an electron band. 
Such local moments arise from electron correlation and 
Hunds coupling 
in the $d$ shells of transition metals or the 
$f$ shells of rare earths. 
Although historically the `Kondo lattice'
arose as the lattice
version \cite{hewson} of the Kondo impurity problem, and refers to 
$S=1/2$  moments coupled to  conduction electrons,
there are also systems with local electron-spin coupling
where the moment is due to a spin with $2S \gg 1$.
In that case the quantum fluctuations of the local moment, and
the Kondo effect itself, are not relevant. 
Such a system can be described by  
a classical Kondo lattice model (CKLM).
This limit is relevant for a wide variety of materials,
{\it e.g}, the manganites \cite{mang-ref}, where
$S=3/2$  moments couple to itinerant electrons via
Hunds coupling, or
$4f$ metals \cite{gd-ref,f-el-gen1,f-el-gen2,f-el-gen3}, {\it e.g},  
Gd with $S = 7/2$,
or the Mn based dilute magnetic semiconductors \cite{dms-ref}
where $S=5/2$. 
In some of these materials, notably the manganites and the magnetic
semiconductors, the coupling scale is known to be large, while in
the $f$ metals they have been traditionally treated as being weak.

The CKLM involves the ordering of `classical' spins,
but the effective interaction between spins  
is mediated by electron delocalisation 
and cannot be described by a short range model.
In fact the major theoretical difficulty in analysing these
systems is the absence of any simple classical spin model. 
Nevertheless, there are two limits where the
CKLM is well understood. (a).~When the electron-spin coupling 
is small, one can perturbatively 
`integrate out' the electrons and
obtain the celebrated  
Ruderman-Kittel-Kasuya-Yosida (RKKY) model \cite{rkky-ref}. 
The effective
spin-spin interaction in this limit is oscillatory and
long range, 
controlled by the free electron susceptibility,
$\chi_0({\bf q})$, and the magnetic ground state is generally
a spiral.
(b).~When the electron-spin coupling is very large compared
to the kinetic energy,  the `double exchange' (DE)  limit, 
the electron spin is 
`slaved' to the orientation of the core spin and the
electronic 
energy is minimised by a ferromagnetic (FM) background \cite{de-old}. 
This leads  intuitively to a spin polarised ground state.

In many materials the ratio of coupling to hopping scale
is $ \ge 1$, but not quite in the double exchange limit.
In that case
one has to solve the coupled spin-fermion model from
first principles. Doing so, particularly in three dimensions
and at finite temperature, has been a challenge.
We study this problem using a combination of
variational calculation and full spin-fermion Monte Carlo.

Our principal results are the following: (i)~We
are able to map out the magnetic ground state all the way 
from the RKKY limit to double exchange, revealing
 the intricate evolution with coupling strength.
(ii)~We demonstrate that the phase boundaries depend
sensitively on electronic hopping parameters. This is
not surprising in the RKKY regime, but the dependence at
stronger coupling is unknown. (iii)~We use our results to
revisit the classic $4f$ magnets,  widely modelled
as RKKY systems, and suggest that with increasing $4f$ 
moment, the effective coupling in these systems pushes
them beyond the RKKY regime. We work out the 
signatures of this `physics beyond RKKY'.

{\it Model:} The Kondo lattice model is given by
\begin{equation}
H = - \sum_{{\langle ij \rangle}  \sigma} t_{ij} 
 c^{\dagger}_{i \sigma}c_{j  \sigma}  -\mu\sum_i n_i
- J\sum_i {\vec \sigma}_i.{\bf S}_i 
\end{equation}
We will use $t=1$ as the  nearest neighbour hopping amplitude, 
and explore a range of $t'$, the next neighbour hopping, on
a cubic lattice.
Changing $t'$ will allow us to explore changes in the (bare)
Fermi surface, and particle-hole asymmetry.  
$\mu$ is the chemical potential, and $J > 0$
is the local electron-spin coupling. We assume the
${\bf S}_i$ to be classical unit vectors, and absorb
the magnitude of the core spin into $J$ wherever
necessary. 
${\vec \sigma}_i$ is the electron spin operator.
We work with $\mu$, rather than electron 
density $(n)$, as the control variable  so that regimes of
phase separation (PS) can be detected, and study the magnetic
properties for varying $n$, $t'/t$, $J/t$, and temperature
$T/t$.

Although there have been many studies in the `double exchange'
($J/t \rightarrow \infty$) limit \cite{de-ref}, 
the attempts to explore the full  $n-J-T$ 
phase diagram have been limited.
$(a)$~An effective action 
obtained from the CKLM via gradient expansion
\cite{grad-exp}
has been analysed. This mapped out some of the 
commensurate and spiral phases 
in two dimensions, where the phases are fewer. 
It  did not explore the finite temperature physics, 
{\it e.g}, the $T_c$ scales, and seems to be inaccurate
when handling commensurability effects near $n=1$.
$(b)$~The model has been studied within dynamical
mean field theory \cite{dmft}
(DMFT), and the broad regimes of ferromagnetism,
antiferromagnetism (AFM), and incommensurate order have  been mapped out.
Unfortunately the effective `single site' character of DMFT 
does not allow a characterisation of the incommensurate phases
and misses out on the richness of the phase diagram.
The loss of information about spatial fluctuations
also means that critical properties, either in magnetism or
transport, cannot be correctly captured. 
$(c)$~An `equation of motion' approach \cite{eom-kien}
has been employed to 
study general {\it finite S} spins coupled to fermions,
and results have been obtained in the classical limit as 
well. However, except the ferro and antiferromagnetic phases
other magnetic states do not seem to have been explored.
$(d)$~The full spin-fermion Monte Carlo, using exact diagnolisation,
has been employed \cite{klm-mc}
in one and two dimensions but severe size limitations
prevent access to non trivial ordered states.

{\it Method:} 
The problem is technically difficult because it 
involves coupled quantum and classical
degrees of freedom, and 
there is in general no equivalent  classical
spin Hamiltonian.  The probability distribution for
spin configurations is given by
$ P\{ {\bf S} \} \propto  Tr_{c, c^{\dagger}} e^{-\beta H} $
so  the  `effective Hamiltonian' is
$ H_{eff}\{ {\bf S} \} =
-{1 \over {\beta}}  log~Tr_{c, c^{\dagger}} e^{- \beta H}$, 
the fermion free energy in an arbitrary 
background $\{ {\bf S}_i \} $. It cannot be 
analytically calculated except when $J/t \ll 1$.

When $J/t \ll 1$, the (free) energy  calculated
perturbatively  
to ${\cal O}(J^2)$  leads to the RKKY spin Hamiltonian \cite{rkky-ref}, 
$H^{eff}_{RKKY} = \sum_{ij} J_{ij} {\bf S}_i. {\bf S}_j$,
where $J_{ij} \sim  J^2 \chi^0_{ij} $ and  $\chi^0_{ij}$ is
the non local susceptibility of the free $(J=0)$ electron
system. $\chi^0_{ij}$ is long range and oscillatory. 
We will analyse this
model to understand the weak coupling phases.
At strong coupling, $J/t \rightarrow \infty$,
there is no exact analytic $H_{eff}$ but we can
construct approximate
self consistent models \cite{sk-pm-scr} of the form
$H^{eff}_{DE} = -\sum_{\langle ij \rangle} D_{ij} \sqrt{1 + {\bf S}_i.
{\bf S}_j } $, with the $D_{ij}$  related to the electronic
kinetic energy.
Unfortunately, when $J \sim {\cal O}(t)$ 
neither the RKKY model nor the DE approximation
are valid. 
This regime requires new tools and we will use a combination of
(i)~variational calculation (VC)  \cite{vc-ref}
for the magnetic ground state, and 
(ii)~spin-fermion Monte Carlo using a `travelling cluster'
approximation \cite{tca-ref} (TCA-MC) at finite temperature.

For the variational calculation we
choose a simple parametrisation \cite{vc-conf}
for the spin configuration:
$ S_{iz} = \alpha $, 
$ S_{ix} = \sqrt{1 - \alpha^2}~ cos  {{\bf q}.{\bf r}_i}$
and 
$ S_{iy} = \sqrt{1 - \alpha^2}~ sin {{\bf q}.{\bf r}_i} $.
This encompasses the standard ferromagnet
and antiferromagnet, as well as planar spiral phases, canted
ferromagnets, and A and C type antiferromagnets.
For a fixed $\mu$ and $J$ we compute the electronic energy 
${\cal E}(\alpha, {\bf q}, \mu)$ and  minimise it with respect to 
$\alpha$ and ${\bf q}$. The electronic density
at the chosen $\mu$ is computed 
on the minimised state. Since the magnetic background only 
mixes electronic states
$\vert {\bf k}, \uparrow \rangle$ and
$\vert {\bf k} - {\bf q}, \downarrow\rangle$ the 
electronic eigenvalues  $\epsilon^{\pm}({\bf k}, {\bf q})$
are simple, and  only an
elementary  numerical sum is required
to calculate ${\cal E}(\alpha, {\bf q}, \mu) = \sum_{{\bf k}, \pm}
\epsilon^{\pm}({\bf k}, {\bf q}) 
\theta(\mu - \epsilon^{\pm}({\bf k}, {\bf q})) $.

While the VC provides a feeling for the possible ground states,
it has the limitation that
(i)~it samples only one family of
(periodic) functions in arriving at the ground state, and
(ii)~finite temperature properties,
{\it e.g}, the magnetisation and the critical temperature
are not accessible.
For this we `anneal' the system towards the  equilibrium
distribution 
$ P\{ {\bf S} \} \propto  Tr_{c, c^{\dagger}} e^{-\beta H} $
using the TCA based Monte Carlo. 
In this method the 
acceptance of a spin update
is determined by diagonalising a cluster Hamiltonian
constructed around the update site, and avoids iterative
diagonalisation \cite{klm-mc} of the  full system. We can access
system size $\sim 10^3$  using a moving cluster of size $4^3$.

The TCA captures phases with commensurate wavevector 
${\bf Q}$ 
quite accurately, but 
access to the weak coupling incommensurate phases is poor.
To get an impression of the ordering temperature for these phases
we compute the energy difference
$\Delta {\cal E}(n, J) = {1 \over N} ({\cal E}_{disord}
-{\cal E}_{ord})$,  between the ordered state
and a fully spin disordered state in a large system. ${\cal E}_{ord}$
is calculated from the variational ground state, and ${\cal E}_{disord}$
by diagonalising the electron system in a fully spin disordered
background on a large lattice. $\Delta {\cal E}(n, J)$ 
is the `condensation energy' of the ordered state, and 
provides a crude measure of the effective exchange and $T_c$.
Where we could compare the trend to MC data, the agreement was
reasonable.

{\it Results:}
The results of the variational calculation in the `symmetric'
($t'=0$) case are shown in Fig.1.
We employed a 
grid with upto $40^3$ ${\bf k}$ points, 
and have checked stability with respect to grid size.  
Let us analyse the weak and
strong coupling regimes first before getting to the more complex
intermediate coupling regime. 

\begin{figure}[t]
\centerline{
\includegraphics[width=7.5cm,height=6.0cm,angle=0,clip=true]{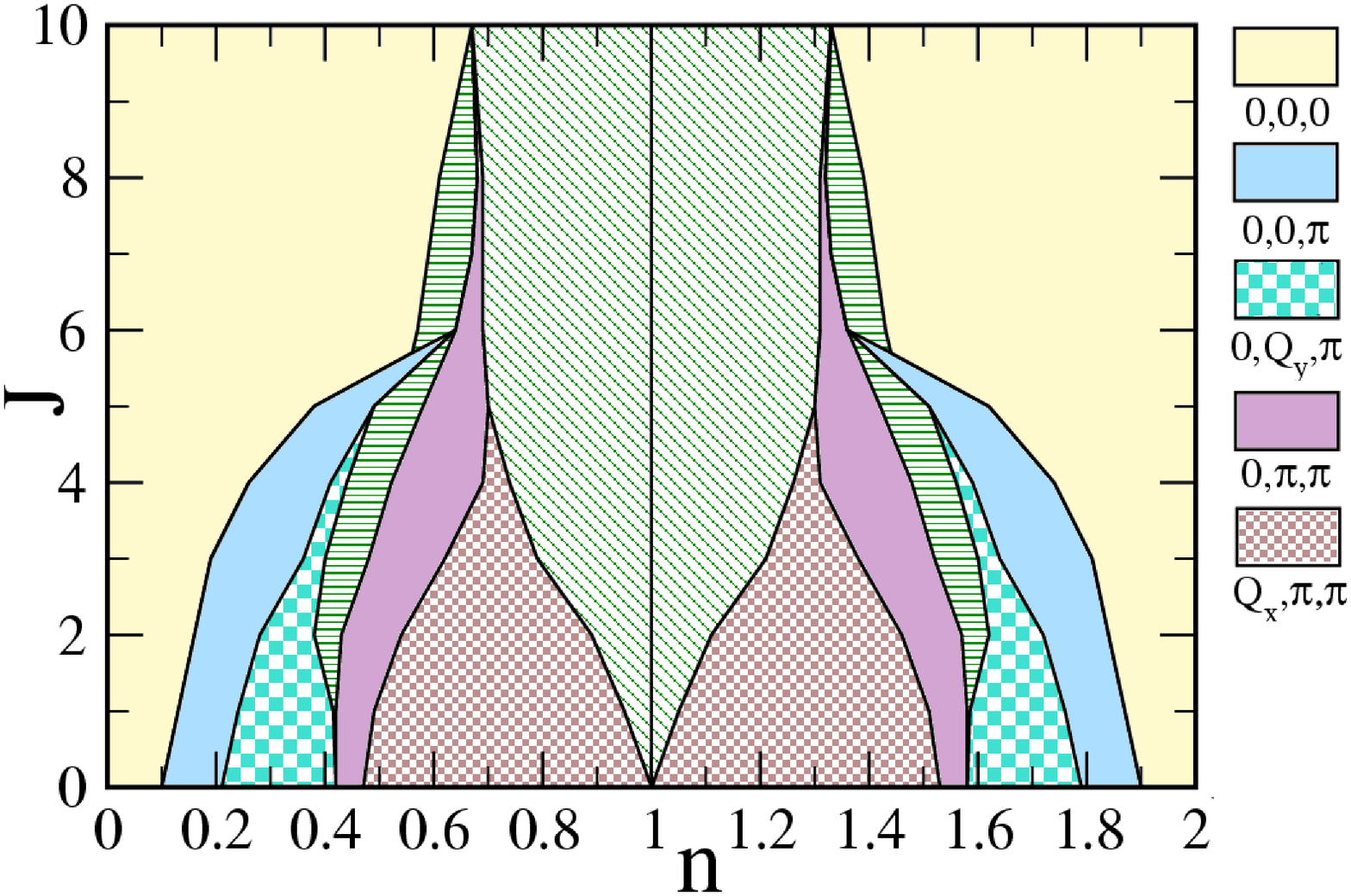}
}
\vspace{.2cm}
\caption{Colour online:
Magnetic ground state for the particle-hole symmetric model 
$(t'=0)$ for varying electron density $(n)$ and electron-spin 
coupling $(J)$.
The phases are characterised by their ordering wavevector
${\bf Q}$, indicated by the colour code in the legend to
the right, and their net magnetisation $\alpha$ (if any).
Among the `commensurate' phases,
${\bf Q} = \{0,0,0\}$ is the usual ferromagnet,
$\{0,\pi,\pi\}$ and $\{\pi,\pi,\pi\}$ 
are antiferromagnets with no net magnetisation, while 
the $\{0, 0, \pi\}$ antiferromagnet has $\alpha=0$ 
for $J \rightarrow 0$ but 
picks up finite magnetisation with increasing $J$.
At $n=1$ the system is always a 
${\bf Q} =\{\pi, \pi, \pi\}$ antiferromagnet.
The incommensurate phases have ordering wavevectors 
$\{Q_x,Q_y,Q_z\}$
of which at least one component is neither $0$ nor $\pi$.
For such phases the exact wavevector depends on the 
value of $n$ and $J$. 
For example, for $J \rightarrow 0$ the (blue) checkerboard region
in the left hand corner, to the right of ${\bf Q} =\{0,0,\pi\}$,
has wavevector ${\bf Q}=\{0,Q_y,\pi\}$, where $Q_y$ varies
from $0$ to $\pi$ as one moves left to right.
The (green)
shaded regions in the phase diagram, {\it not indicated} in the
legend, are windows of  phase separation.
No homogeneous phases are allowed in these regions.
The results in this  figure are based on a variational
calculation using a $20^3$ ${\bf k}$ point grid, 
and cross-checked with data on  $40^3$.
}
\end{figure}

{\bf (i)}~{\it RKKY limit:}
The key features for $J/t \rightarrow 0$ are:
(i)~the occurence of `commensurate' planar spiral 
phases, with wavenumber 
${\bf Q}$ which is $\{0,0,0\}$, or
$\{0, 0, \pi\}$, {\it etc}, over  {\it finite} density windows, 
(ii)~the presence of planar 
spirals with incommensurate
${\bf Q}$ over certain density intervals, (iii)~the absence of any
phase separation, {\it i.e}, only second order  phase boundaries, and
(iv)~the presence of a `G type', ${\bf Q} =\{\pi, \pi, \pi\}$,
antiferromagnet at $n=1$.
Although the magnetic state is obtained from the variational
calculation, much insight
can be gained by analysing the $H^{eff}_{RKKY}$. Since the spin-spin
interaction is long range it is useful to study the Fourier 
transformed version $H^{eff}_{RKKY} \equiv \sum_{\bf q} {\tilde J}_{\bf q}
\vert {\bf S}_{\bf q} \vert^2$, where ${\tilde J}_{\bf q}
= \sum_{i-j} J_{ij} e^{i {\bf q}.{\bf R}_{ij}}$ 
and 
${\bf S}_{\bf q} = \sum_{i} {\bf S}_i e^{i {\bf q}.{\bf R}_{i}}$.
The coupling ${\tilde J}_{\bf q} = J^2 \chi_0({\bf q}, n)$ is controlled
by the spin susceptibility, $\chi_0({\bf q}, n)$,
of the $J=0$ 
tight binding electron system. For our choice of
variational state 
the minimum of $H^{eff}$ corresponds to
the wavevector at which $\chi_0({\bf q}, n)$ has a maximum.
We independently computed $\chi_0({\bf q}, n)$ and confirmed 
\cite{kp-pm-unpub} that
the wavevector ${\bf Q}(n)$ obtained from the VC closely matches
the wavevector ${\bf q}_{max}(n)$ of 
the peak in $\chi_0({\bf q}, n)$.
The absolute maximum in 
$\chi_0({\bf q}, n)$ remains at ${\bf q} = \{0, 0, 0\}$, as the
electron density is increased from $n=0$, and at a critical
density ${\bf q}_{max}$ shifts to $\{0, 0, \pi\}$.
With further increase in density ${\bf q}_{max}$ evolves
through $\{0, q, \pi\}$ to the C type $\{0, \pi, \pi\}$,
then $\{q, \pi,  \pi\}$, and finally the G type AFM with
$\{\pi, \pi,  \pi\}$, where the Fermi surface is nested.
The absence of `conical' phases, with finite $(\alpha)$ and a spiral
wavevector, is consistent with what is known in the RKKY problem.

\begin{figure}
\centerline{
\includegraphics[width=8.5cm,height=8.0cm,angle=0,clip=true]{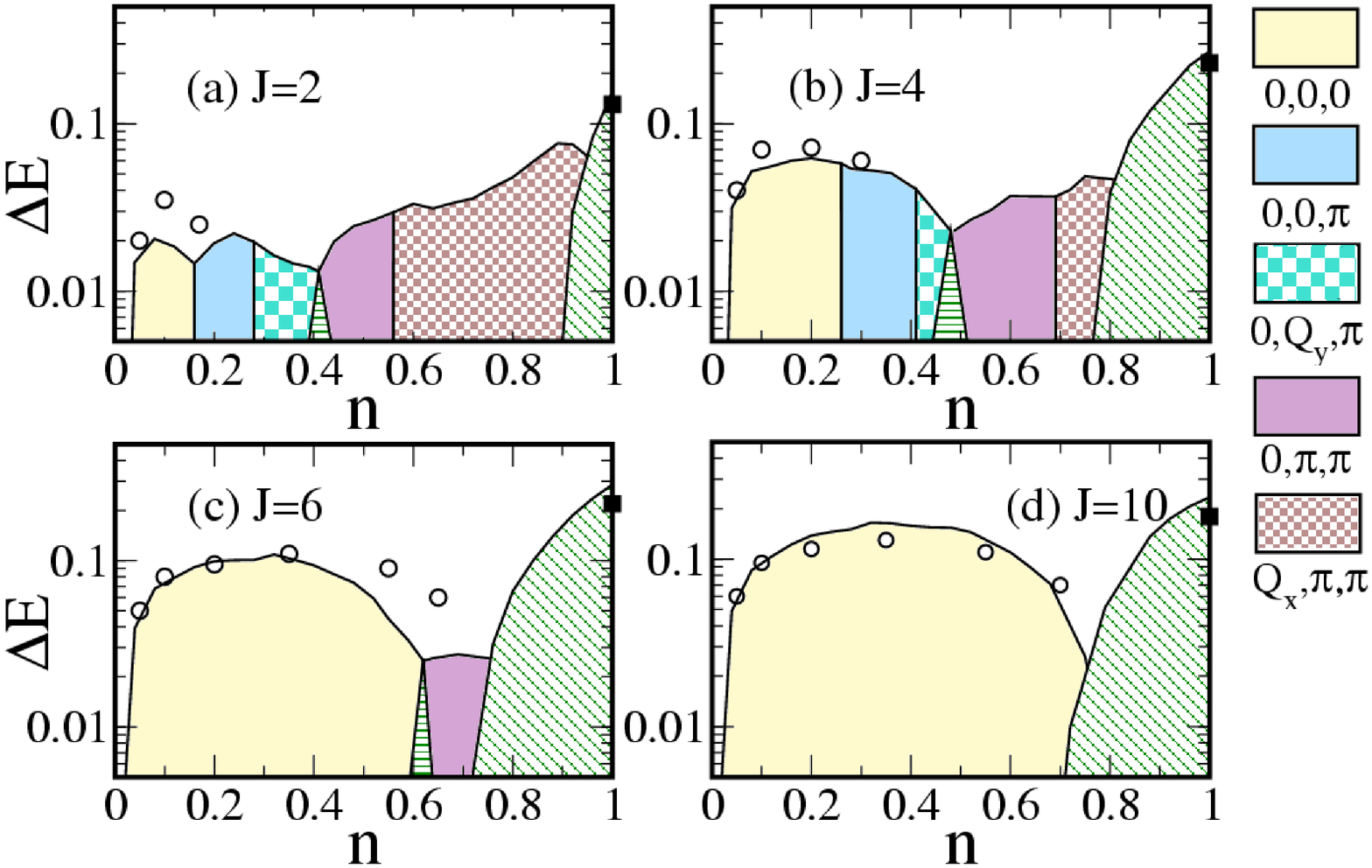}}
\vspace{.2cm}
\caption{Colour online:
The finite temperature phase diagram in the
particle-hole symmetric case, for various $J$.
Panels (a)-(d) show the different ordered phases 
and their estimated
transition temperature as we move from the
weak coupling to the double exchange limit. 
The legend for the phases is shown on the right.
The transition temperatures
are  based either on Monte Carlo results (shown as symbols),
or the $\Delta {\cal E}$ estimate (firm lines) described in
the text.
Notice that the $T_c$ for the ferromagnetic,
${\bf Q} =\{0,0,0\}$, phase increases (and saturates) with
increasing $J$. At $n=1$ the order is at
${\bf Q} =\{\pi,\pi,\pi\}$
and the corresponding $T_c$ initially
increases with increasing $J$ and then decreases. 
Except for ${\bf Q}= \{0,0,0\}$ and $\{\pi,\pi,\pi\}$
other phases vanish
by the time $J/t=10$. The Monte Carlo estimate of 
ferromagnetic $T_c$
are shown as circles, while that of the antiferromagnet is
marked on the $n=1$ axis by a square symbol.
As in Fig.1 the (green) shaded regions indicate phase separation.
}
\end{figure}

There is no phase separation, {\it i.e}, discontinuities in
$n(\mu)$, for $J/t \rightarrow 0$ since
the $\mu-n$ relation is that of the underlying
tight binding system and free of any singularity. The
phase transitions with changing $n$ are all
{\it second order}.  With
growing $J/t$, however, some phase boundaries become
first order and  regimes of PS will emerge.

{\bf (ii)}~{\it Strong coupling:}
For $J/t \rightarrow \infty$, it makes sense to quantise the
fermion spin at site ${\bf R}_i$ 
in the direction of the core spin ${\bf S}_i$, and project
out the `high energy' unfavourable state. This leads to 
an effective  spinless  fermion problem whose bandwidth 
is controlled by the  average spin overlap 
$\langle {\bf S}_i.{\bf S}_j \rangle$ 
between neighbouring sites. 
The overlap is largest for a fully polarised state, and the
FM turns out to be the ground state at all $n \neq 1$.
At $n=1$ `real hopping' is forbidden so the fermions
prefer a G type AFM background to 
gain kinetic energy ${\cal O}(t^2/J)$
via virtual hops.

The FM and G type AFM have a first order
transition between them with a window of phase separation,
easily estimated at large $J/t$.
The fully polarised FM phase has a density of states (DOS) 
which is simply two 3D tight binding DOS with splitting 
$J$ between the band centers. If we denote this DOS as
$N_{FM}(\omega,J)$ then the energy of the FM phase
is ${\cal E}_{FM}(\mu, J) = \int_{-\infty}^{\mu} N_{FM}(\omega,J) \omega 
d\omega$, and the particle 
density is $n(\mu, J) = \int_{-\infty}^{\mu} N_{FM}(\omega,J) 
d\omega$.
There will be corresponding expressions when we consider 
electrons in the $\{\pi, \pi, \pi\}$ AFM background,
with DOS $N_{AFM}(\omega,J)$.
Once
we know $\mu = \mu^{FM}_{AFM}$ that satisfies ${\cal E}_{FM}(\mu, J) = 
{\cal E}_{AFM}(\mu, J)$ we can determine the PS window from the
density equations. 
Since the FM phase has a dispersion $\epsilon^{FM}_{\bf k}
= \epsilon_{0,{\bf k}} \pm J/2$,
where $\epsilon_{0,{\bf k}} = -2t(cosk_x a + cosk_y a + cosk_z a)$,
and the AFM phase has dispersion $\epsilon^{AFM}_{\bf k}
= \pm \sqrt{\epsilon_{0,{\bf k}}^2 + (J/2)^2}$, it is
elementary to work out  $\mu^{FM}_{AFM}$.
The analysis can be extended to several competing phases.
It is significant that even at $J/t =10$, which might
occur for strong Hunds' coupling in some materials, 
the FM phase occurs only between $n=\{0,0.7 \}$. 

\begin{figure}
\centerline{
\includegraphics[width=6.0cm,height=5.0cm,angle=0,clip=true]{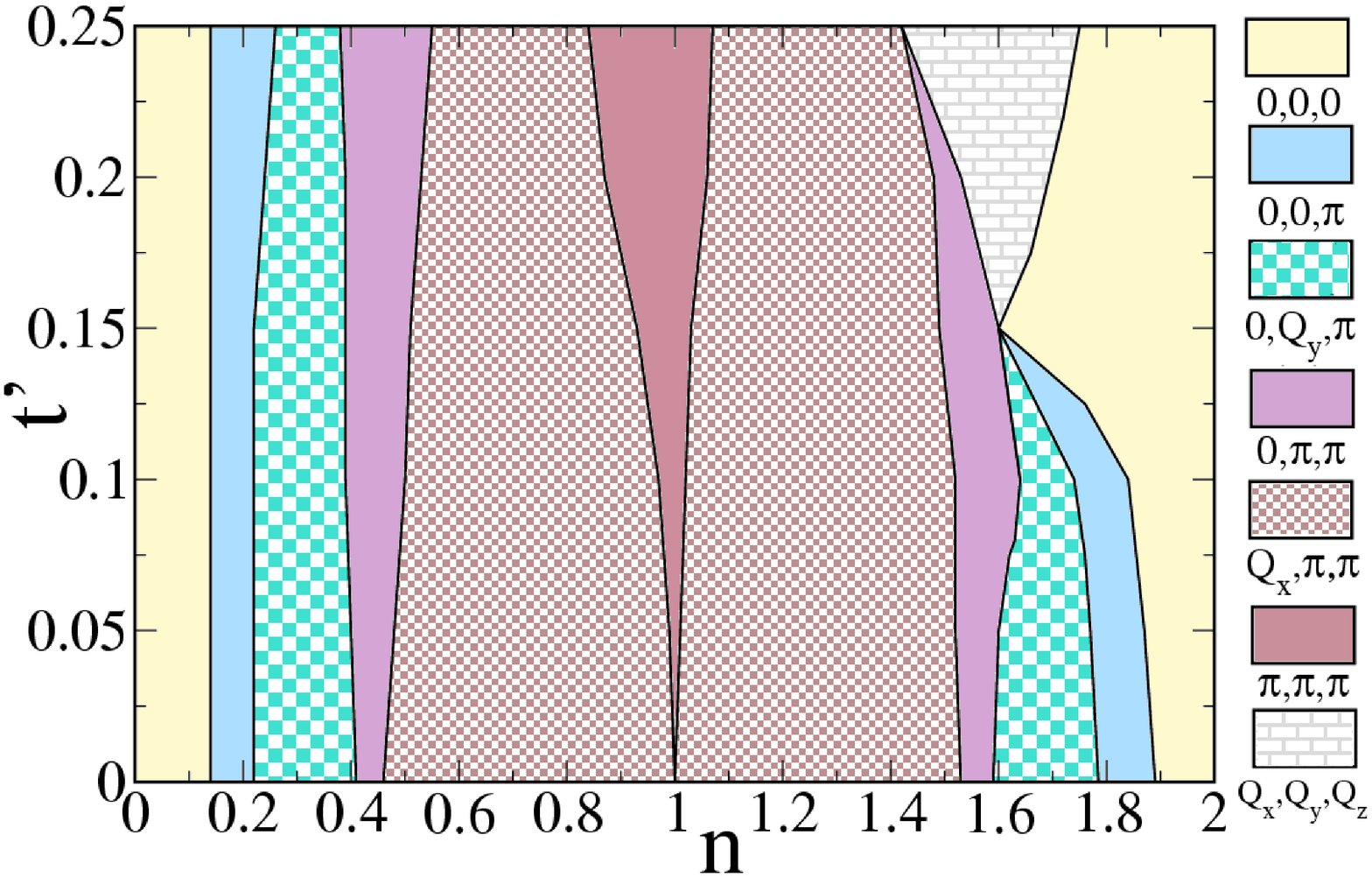}}
\caption{Colour online: The magnetic ground state in the 
RKKY limit, showing the dependence of the ordering wavevector
${\bf Q}$ on electron density and particle-hole asymmetry (via $t'$).
The legend for the various states is shown on the right.
The calculations were done at weak coupling, $J = 0.5$. 
Note the growing asymmetry of the phases (about $n=1$)
as $t'$ increases. It is also clear that if the
hopping parameter $t'$ changes (due to pressure, {\it etc})
the magnetic ground state can change even if the electron
density remains fixed, as discussed for $4f$ systems
in \cite{f-nat-pap}. This is particularly prominent
in the top right hand corner of the figure.
In constructing this phase diagram we
have ignored a narrow sliver of
phase separation near $n=1$. 
}
\end{figure}

{\bf (iii)}~{\it Intermediate coupling:}
The intermediate coupling regime is where one is outside the
RKKY window, but not so large a coupling that only the
FM and G type AFM are possible. Towards the weak coupling end it
implies that the planar spirals begin to pick up a net
magnetisation, $\alpha$, and now become `conical' phases.
 Windows of phase separation also appear, 
particularly prominent between the $\{ Q_x, \pi, \pi\}$ and
G type AFM (near $n=1$), and suggest the possibility of
inhomogeneous states, {\it etc}, in the presence of
disorder.  
The prime signature of `physics beyond RKKY',
however, is that the RKKY planar spirals
now pick up a net magnetisation and
much of the phase diagram starts to evolve towards the
ferromagnetic state. 

{\bf (iv)}~{\it Finite temperature:}
The TCA based MC readily captures the FM and $\{\pi, \pi, \pi\}$
AFM phases at all coupling. However, it 
has difficulty in capturing the
more complex spiral, A, and C type phases when we `cool' from the
paramagnetic phase. In the intermediate $J$ regime it usually yields
a `glassy' phase with the structure factor having weight distributed
over all ${\bf q}$. In our undertstanding this is a limitation of
the small cluster based TCA, and the energies yielded by VC are
better than that of `unordered' states obtained via MC.
To get a feel for the ordering temperature we have calculated
the energy difference $\Delta {\cal E}$, defined earlier,
as often done in electronic structure calculations. This
provides the trend in $T_c$ across the phases, Fig.2, and
wherever possible we have included data about actual $T_c$
(symbols) obtained from the MC calculation.
Broadly, with increasing $J$  the $\Delta {\cal E}$ and 
$T_c$ scales increase
but the number of phases decrease.
The $T_c$ of the G type AFM is expected to fall at large
$J$ but even at $J/t=10$ it is larger than the peak
FM $T_c$.

\begin{figure}[t]
\centerline{
\includegraphics[width=7.0cm,height=12.0cm,angle=0,clip=true]{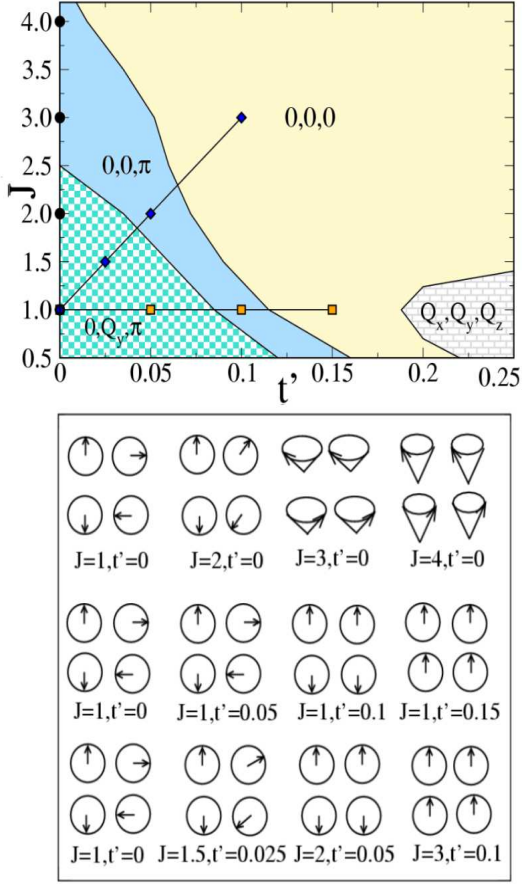}}
\caption{Colour online: {\bf Top:} the magnetic ground state at $n=1.7$
for varying $t'$ and $J$. The ordering wavevector is
marked on the phases. 
The magnetic order has a pronounced dependence on {\it both}
the `bandstructure' (through $t'$)
 and the electron-spin coupling.
We highlight three kinds of parameter variation. 
(i)~Varying $J$
at fixed $t'$, the points on the $y$ axis,
shows how changing electron-spin coupling can change
the ground state. 
(ii)~Varying $t'$ at weak coupling, $J=1$, illustrates
how bandstructure affects the RKKY magnetic order. 
(iii)~In the $4f$ elements we think what happens is
a combination of (i) and (ii) above, as shown by points on 
the diagonal. 
{\bf Bottom:} An impression of the real space spin configuration
for the three parameter sets (i)-(iii) in the top panel.
Each $2 \times 2$ pattern is for a $t',~J$ combination.
The bottom left spin in each pattern is set on the
reference site ${\bf R} =\{0, 0\}$, say. The neighbouring 
three
spins are at $\{{\hat x}, 0\}$, $\{0, {\hat y}\}$, and
at $\{{\hat x}, {\hat y} \}$, where
${\hat x}$ and  ${\hat y}$ are unit vectors on the
lattice. There is no variation in the $z$ direction 
so we only show the in-plane pattern.
Top row: scan (i) above, changing $J$ at $t'=0$. 
Middle row: scan (ii),  changing
$t'$ at $J=1$. Bottom row: scan (iii), 
simultaneous change in $t'$ and $J$.
}
\end{figure}

{\bf (v)}~{\it Interplay of FS and coupling effects:}
Till now we have looked at the particle-hole symmetric case
where $t'=0$. The tight-binding parametrisation of the {\it ab 
initio} electronic structure of any material usually requires
a finite $t'$, in  addition, possibly, to multiple bands.
We will use the $t-t'$ parametrisation of band structure
due to its simplicity. It will also allow us 
to mimic the physics in the $4f$ metals.

At weak coupling the magnetic order is controlled as usual
by the band susceptibility, $\chi_0({\bf q},n)$ which, now,
also depends on 
$t'$. At fixed $n$ the magnetic order
can change simply due to changes in the underlying electronic
structure. 
Our Fig.3 illustrates this dependence, where we use
$J=0.5$ to stay in the RKKY regime and explore the 
variation of magnetic order with $n$ and $t'$. The
range of $t'$ variation is modest, $\sim \{0-0.3\}$,
but can lead to phase changes (at fixed $n$) in some
density windows. We have cross checked the phases with
the peak in $\chi_0({\bf q})$. 

A complicated and more realistic version of this
has been demonstrated recently \cite{f-nat-pap} in the $4f$
family for the heavy rare earths from 
Gd to Tm. 
These elements all have the same hcp crystal structure, and
the same conduction electron count,  
$5d^16s^2$, so nominally the same band filling. 
However, the
electronic structure and Fermi surface changes due
to variation in the lattice parameters and unit cell
volume (lanthanide contraction) 
across the series. It has been argued \cite{f-nat-pap}
that this changes the location
${\bf q}_{max}$ of the peak in $\chi_0$, and 
explains the change in magnetic order from planar
spiral (in Tm) to  ferromagnetism in Gd. A similar 
effect is visible in our Fig.3 where at $n=1.7$, say,
the ordering wavevector changes
from a spiral to FM as $t'$ changes from zero to $0.15$.
In this scenario, $J$ does not affect the
magnetic order but merely sets the scale for $T_c$.
The RKKY interaction strength scales as 
$J_{eff}^2 \sim J^2S(S+1)$, and a similar scaling of the
experimentally measured  $T_c$ is taken as 
`confirmation' 
of the RKKY picture.

Should'nt we also worry about the effect of the
growing $J_{eff}(S)$ on the {\it magnetic order itself}?
If the maximum $J_{eff}$, for Gd with $S=7/2$, were
smaller than the effective hopping scale $t$, then
we need not - the  RKKY scheme would be valid for the
entire $4f$ family. 
However, measurements and electronic structure 
calculations \cite{gd-ref}
in Gd suggest that $J \sim 0.3$eV and
$J_{eff}(7/2) \sim 1$eV. 
The effective $t$ is more ambiguous, since there are multiple
bands crossing the Fermi level, but the typical value is
$\sim 0.3$eV.
This suggests $J_{eff}/t \sim 3$, clearly outside the RKKY window!
What is the consequence for magnetic order, and physical properties
as a whole?

Fig.4 shows the $t'-J$ magnetic phase diagram at $T=0$ for
$n=1.7$. At $t'=0$, the {\it vertical} scan, changing $J$ reveals how
the ordered state changes with increasing $J$ even with electronic
parameters (and hence $\chi_0$ and FS) fixed. 
We have already seen this in Fig.1
The spirit of RKKY is to assume $J \rightarrow 0$, 
and move {\it horizontally}, changing $t'$ across the
series so that one evolves from a planar spiral to a ferromagnet.
We suggest that in the $f$ metals, the parameter points 
are actually on a `diagonal', with increasing $t'$ 
(our version of changing electronic structure) being 
accompanied by increase in $J_{eff}$. 
To capture the trend we set, 
$t'=0$ and $J_{eff} =1.0$ for $S=1$, where the
system is known to be a spiral, and $t'=0.1$ and $J_{eff}=3.0$ for
$S=7/2$ (the case of Gd), and explore the linear variation
shown in Fig.4. This parametrisation is only meant to highlight
the qualitative effect of changing electronic structure and
$J_{eff}$ and since real $t'$ values, {\it etc}, would
need to be calculated from an {\it ab initio} solution.

Within this framework, while the small $S$ result is same
for both RKKY and explicit inclusion of $J_{eff}$, the order
obtained at intermediate $S$ depends on whether one ignores
$J_{eff}$ (as in RKKY) or retains its effect. For
a given $t'$ the phase on the diagonal is quite different
from the phase on the horizontal line.

In fact there is evidence from earlier {\it ab initio} calculations
\cite{nord-mav} that in addition to unit cell volume and $c/a$
ratio, {\it the strength of the $4f$ moment} (and so
$J_{eff}$) also affects the magnetic order. As an illustrative case,
the optimal  spiral wavevector in Ho evolves towards ${\bf Q}
= \{0, 0, 0\}$ as the effective moment is (artificially) varied
from $2\mu_B$ to $4\mu_B$ (Fig.2 in Nordstrom and Mavromaras \cite{nord-mav}). 
If magnetism in this element, and
the $4f$ family in general, were completely determined by
RKKY there would be {\it no dependence on $J_{eff}$}. 
In fact the authors suggested that one should re-examine the 
basic assumptions of the `standard model' of $4f$ magnetism 
\cite{f-el-gen2},
which gives primacy to the RKKY interaction (and magnetoelastic
effects) since the {\it ab initio} results suggest a role
for the effective exchange in the magnetic order.
Our aim here has been to clarify the physics underlying such
an effect within a minimal model Hamiltonian.
This approach would be useful to handle non collinear
phases in complex many band systems,
{\it without any weak coupling assumption}, 
once a tight binding parametrisation of the electronic
structure is available

Let us conclude. We have examined the Kondo lattice model with
large $S$ spins and established the ground state all the way
from the RKKY regime to the strong coupling limit. The intermediate
coupling window reveals a competition between RKKY effects, which
tend to generate a planar spiral, and the tendency to gain exchange
energy via ferromagnetic polarisation. This generally leads to
a `conical' helix, giving way at strong coupling to the
double exchange ferromagnet. Using these results we re-visited
the classic $4f$ magnets to demonstrate how the magnetic phases
there are probably controlled non RKKY spin-fermion effects.
One can add anisotropies and magneto-elastic couplings
to our model to construct a more comprehensive description
of $4f$ magnetism.

\vspace{.2cm}

We acknowledge use of the Beowulf cluster at HRI, and thank Sanjeev Kumar and
B. P. Sekhar for collaboration on an earlier version of this problem.


\begin{thebibliography}{99}
\bibitem{vc-conf} 
Our choice corresponds to a periodically varying 
azimuthal angle, $\phi_i$. The most general periodic state involves
periodic variation in the polar angle $\theta_i$ as well \cite{vc-ref}.
\end{thebibliography}
\end{document}